\documentclass[9pt,twocolumn,twoside]{osajnl}

\journal{ol} 
\usepackage{gensymb}
\usepackage{verbatim}
\setboolean{shortarticle}{true} 

\title{Quantum Airy Photons}

\author[1,2,$\dagger$]{Stephanie Maruca}
\author[1,2,$\dagger$]{Santosh Kumar}
\author[1,2]{Yong Meng Sua}
\author[1,2]{Jia-Yang Chen}
\author[1,2,3]{Amin Shahverdi}
\author[1,2,*]{Yu-Ping Huang}
\affil[1]{Department of Physics, Stevens Institute of Technology, Hoboken, NJ, 07030, USA}
\affil[2]{Center for Quantum Science and Engineering, Stevens Institute of Technology, Hoboken, NJ, 07030, USA}
\affil[3]{Department of Electrical and Computer Engineering, Stevens Institute of Technology, Hoboken, NJ, 07030, USA}

\affil[*]{yuping.huang@stevens.edu}
\affil[$\dagger$]{These two authors contributed equally to this work.}
 \dates{Compiled \today}

\ociscodes{(070.6120) Spatial light modulators; (050.1940) Diffraction; (190.0190) Nonlinear optics; (030.5260)   Photon counting; (270.0270) Quantum optics. } 

\begin{abstract}
With exotic propagation properties, optical Airy beams have been well studied for innovative applications in communications, biomedical imaging, micromachining, and so on. Here we extend those studies to the quantum domain, creating quantum correlated photons in finite-energy Airy transverse modes via spontaneous parametric down conversion and sub-sequential spatial light modulation. Through two-photon coincidence measurements, we verify their Airy spatial wavefunctions, propagation along a parabolic trajectory, and that the spatial modulation does not introduce any observable degradation of quantum correlation between the photons. These results suggest the feasibility of using spatially structured photons for practically advantageous quantum applications.  
\end{abstract}

\setboolean{displaycopyright}{true}

\begin{document}

\maketitle

There has been growing interest in surpassing the diffraction limit of light beams while engineering their directionality for advanced imaging and remote sensing of chemical and biological agents \cite{Betzig92,Rust06,Choi11,Liu17}. In this pursuit, Airy wave packets arose as a unique and promising candidate for their non-diffracting propagation along parabolic trajectories that mimic free acceleration \cite{Berry79}. This extraordinary phenomenon was observed more recently by using finite-energy optical Airy beams, together with self healing where their spatial modes are self reconstructed after being partially blocked \cite{Siviloglou07,Christodoulides07,Siviloglou08}. A common way to create the Airy beams is via spatial modulation of Gaussian beams using a cubic phase mask and a Fourier lens \cite{Siviloglou07,Christodoulides07,Siviloglou08}, which allows control of the peak-intensity locations during their ballistic propagation by laterally translating the phase mask or by tilting the Fourier lens \cite{Siviloglou08,Hu10,Hu12}. Alternatively, they can also been created directly in microchip lasers \cite{Longhi11,Porat11} and through nonlinear optical processes \cite{Ellenbogen09}. Besides optical Airy beams, their electronic counterparts have also generated by diffraction of electrons through a nanoscale hologram \cite{Voloch13}.

Thus far, optical Airy beams have been explored for exotic applications \cite{Hu12} such as micro-particle manipulation \cite{Baumgart08,Singh17}, surface plasmonic bending \cite{Salandrino10,Minovich11, Zhang11}, directioned filamentation \cite{Chong10,Abdollahpour10,Panagiotopoulos13}, parabolic plasma channeling \cite{Polynkin09}, micromachining along a curve \cite{Mathis12}, and superresolution imaging \cite{Jia14}. More recently, Airy beams have been exploited in optical parametric oscillators \cite{Aadhi16,Aadhi17} and four-wave mixing in atomic vapor cells \cite{Wei14}, with growing applications in medical science \cite{Vettenburg14}, defense \cite{Extance15}, and optical communications \cite{Gu10,Rose13b,Rose13}. Those exciting studies, however, are all in the classical domain. It remains to examine if those distinct advantages found in Airy beams could be exploited in the quantum domain \cite{KILLINGER, Kirmani14, Maccarone15,Giovannini15,Piksarv17}.

In this paper, we generate quantum correlated photon pairs in finite-energy Airy modes through spontaneous parametric down conversion (SPDC) and subsequentially spatial light modulation using a cubic phase mask followed by a Fourier lens. Through time-correlated photon counting, we verify the Airy spatial wavefunctions of the photons and their propagation along a parabolic trajectory. Importantly, we find that the spatial modulation on the photons does not introduce any observable degradation of quantum correlation between them. Those Airy photons could be deployed in various quantum applications for remote sensing, imaging, communications, etc, where distinct advantages may be expected \cite{Villa12, Tyler09,Nelson14,Yao2006,Etcheverry13,Fickler2013,Patarroyo2013PRL,Chrapkiewicz2016,TaylorBowenPhysRep2016}.

\begin{figure*}[htbp]
   \begin{center}
 \includegraphics[width=5in]{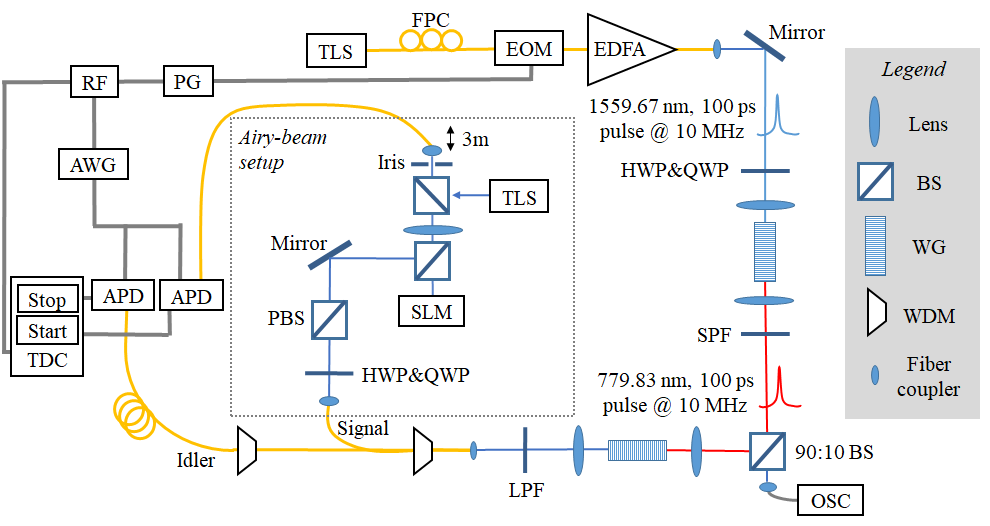} 
   \end{center}
    \caption{Experimental setup for generation and detection of  quantum correlated Airy photons. TLS: Tunable Laser Source, FPC: Fiber Polarization Controller, EOM: Electro-optic Modulator, EDFA: Erbium-doped Fiber Amplifier, QWP: Quarter-Waveplate, HWP: Half-Waveplate, OSC: Oscilloscope, BS: Beamsplitter, PBS: Polarizing Beamsplitter, SLM: Spatial Light Modulator, WG: Waveguide, PG: Pulse Generator, AWG: Arbitrary Waveform Generator, RF: Radio-Frequency Source, APD: InGaAs Avalanche Photodiode, TDC: Time-to-Digital Converter, SPF: Short Pass Filter, LPF: Long Pass Filter, WDM: Wavelength Division Multiplexer.}
    \label{fig1}
 \end{figure*}
 
Figure \ref{fig1} outlines our experimental setup. A continuous-wave (CW) laser (LaserBlade, Coherent Solutions) generates light at 1559.67 nm with $\leq$ 100 kHz linewidth. The light passes through an electro-optic modulator to create a pulse train with 100 ps full width at half maximum (FWHM) and 10 MHz repetition rate, synchronized with a radio-frequency source that gates single photon detectors for measurement. The optical pulses are then amplified in an Erbium-doped Fiber Amplifier to obtain high peak power ($\sim 1$ W), and guided through a fiber collimator, a half-waveplate, and a quarter-waveplate, before coupled into a magnesium-doped periodically poled lithium niobate (PPLN) waveguide through an aspheric lens. The PPLN, about 1 cm long, is temperature stabilized and phase matched to create SPDC pump pulses at 779.83 nm via second harmonic generation. The output pulses are then filtered with three short-pass filters which provide a total $>$180 dB extinction to remove any residual fundamental light. A 90:10 beamsplitter is used to tap 10\% power of the second harmonic light for real-time monitoring and the remaining power is guided through an aspheric lens into the second PPLN waveguide with similar phase matching characteristics for photon-pair generation via SPDC (779.83 nm $\rightarrow$ 1554.7 nm + 1564.7 nm) \cite{Ghosh87,Takesue05,Ma09,Oza14,YuPing14}. The created photon pairs are coupled out and collimated by another aspheric lens before passing through a long pass filter with extinction of 50 dB to remove the 779.83-nm light pump pulses. After coupled into a fiber, the signal and idler photon pairs are picked at 1554.7 nm and 1564.7 nm respectively, using wavelength division multiplexers. The idler photons are guided through an optical fiber delay line and detected using an InGaAs avalanche photodiode (InGaAs APD) (ID210, ID Quantique) with 10\% quantum efficiency and $<100$ dark counts per second. The signal photons are guided through a fiber collimator to the free space Airy-beam setup, as shown in Fig.~\ref{fig1}, before they are detected also using a second InGaAs APD. Each detected photon produces a TTL pulse which goes into a multi-channel time-to-digital converter (SENLS, HRM-TDC) for coincidence measurement.

The Airy-beam setup follows the standard approach of first producing a cubic phase modulation on a Gaussian incident beam using a  1.5 cm $\times$ 1.1 cm spatial light modulator (SLM) (Santec SLM-100), and then passing it through a Fourier transform lens \cite{Christodoulides07,Siviloglou07}. Here, two-dimensional Airy spatial modes are created for both classical light and the signal photons, whose transverse mode is given by
\begin{eqnarray}
E(x,y) = \mathbf{Ai}\Big(\frac{x}{x_0}\Big) \,\mathbf{Ai}\Big(\frac{y}{y_0}\Big) \, \mathrm{exp} \Big[ a\Big( \frac{x}{x_0}+\frac{y}{y_0}\Big) \Big],
\end{eqnarray}
where $\mathbf{Ai}$ stands for the Airy function, $x_0$ and $y_0$ are the scaling factors in $x$ and $y$ transverse directions, respectively, and $a$ is a truncation factor \cite{Polynkin09}. In this experiment, $x_0=y_0=271\, \mu m$ which corresponds to an Airy spatial mode with a main lobe $ \sim 434\, \mu m$ FWHM. 
The phase mask is created numerically by discretizing the cubic phase over an 1.04 cm $\times$ 1.04 cm area (smaller than the SLM screen) with pixel size $\sim$ 10.4 $\mu m$. A linear phase modulation along the $x$ direction is superposed on the cubic phase mask to deflect the modulated light into its first diffraction order \cite{XZWang12} thus separating it from any unmodulated light that will be in the zeroth order. The total loss as the light go through the free-space setup is 22 dB. The spatial profile of the resulting two-dimensional Airy beam is directly measured using a NIR-IR camera (FIND-R-SCOPE Model No. 85700) with pixel resolution of 17.6 $\mu$m. 

Once the Airy-beam setup is verified and optimized using a CW laser at 1568.7 nm, the signal photons are switched in the same fiber and free-space optical paths. At the focal point of the Fourier lens, they are spatially modulated to be in a two-dimensional Airy wavefunction. From there, those Airy photons propagate for 3 meters along a parabolic, self-accelerating trajectory. Both the Airy wavefunction and parabolic trajectory are verified using quantum correlation measurement with the idler photon, by first collecting the Airy photons in a fiber and detecting them using an InGaAs APD. 

\begin{figure}[htbp]
   \begin{center}
   \includegraphics[width=3.4 in]{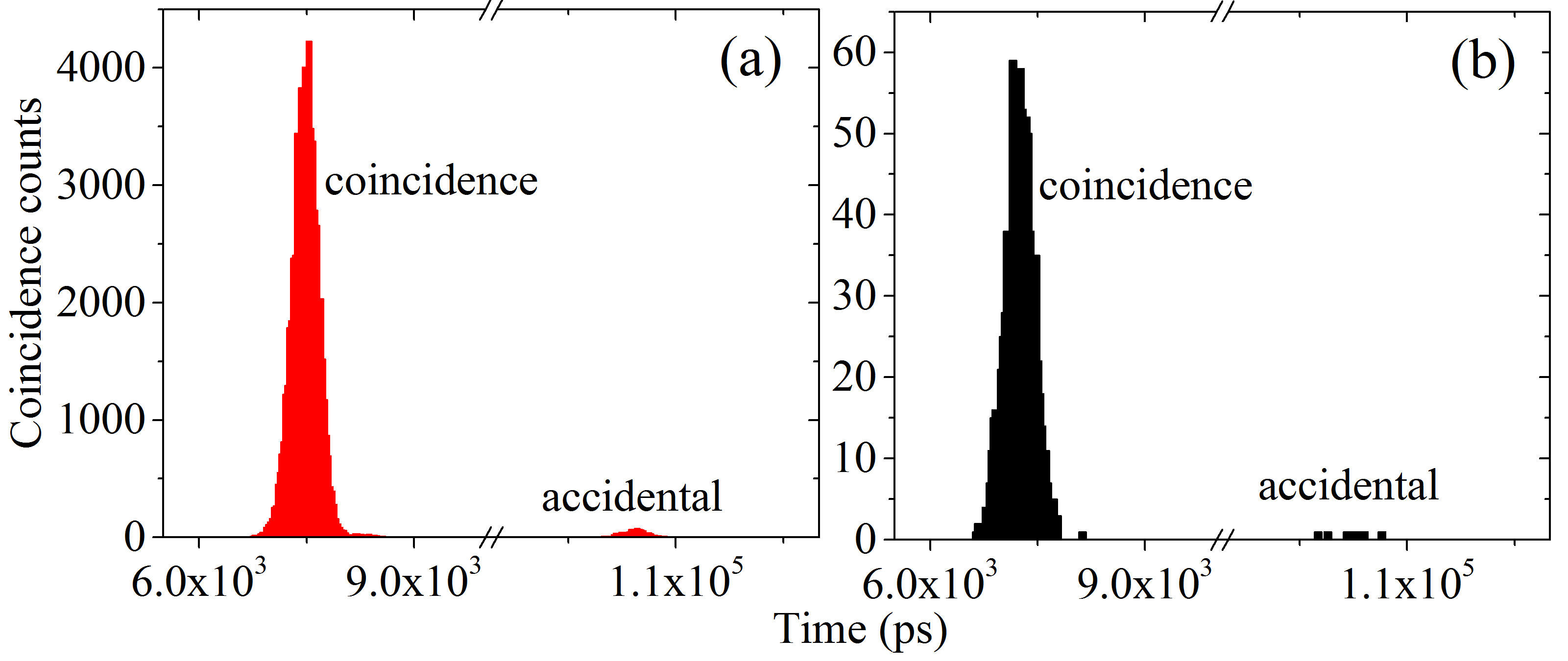}
   \end{center}
    \caption{Two-photon coincident detection for (a) unmodulated photons and (b) Airy photons.}
    \label{fig2}
 \end{figure}

We first measure the quantum correlation between the signal and idler photons with and without the Airy modulation. For the measurement without modulation, the signal photons bypass the Airy-beam setup and propagate instead through an optical fiber of equivalent path length. Using the standard experimental procedures in \cite{QPMS2017}, the coincident-to-accidentals ratio (CAR) of the photon pairs is measured over an integration time of 60 minutes with 33 mW pump pulse peak power. For the measurement with the modulation, the signal photons are guided through the Airy-beam setup and collected by a single-mode fiber (SMF-28) at the focal point of the Fourier lens using a fiber collimator consisting of an aspheric lens (Thorlabs C220TMD-C). The lens is chosen to have a clear aperture that is much larger than the main lobe of the Airy beam and a numerical aperture that matches the fiber for the maximum coupling efficiency. The measurement results are shown in Fig.~\ref{fig2}, where the coincident counts for the unmodulated and Airy signal photons are compared. The CAR values in these two cases are 69$\pm 6$ and 70$\pm 3$, respectively, which are within each's statistical error despite substantial loss in the current Airy-beam setup that reduces the photon counts. This result indicates that the phase modulation by the SLM does not have any measurable impact on the quantum states of the photon pairs \cite{Yao2006}, which is essential for their applications in quantum imaging \cite{Fickler2013}, distant object identification \cite{Patarroyo2013PRL}, quantum key distribution \cite{Etcheverry13}, and so on.
 
\begin{figure}[htbp]
   \begin{center}
 \includegraphics[width=3 in]{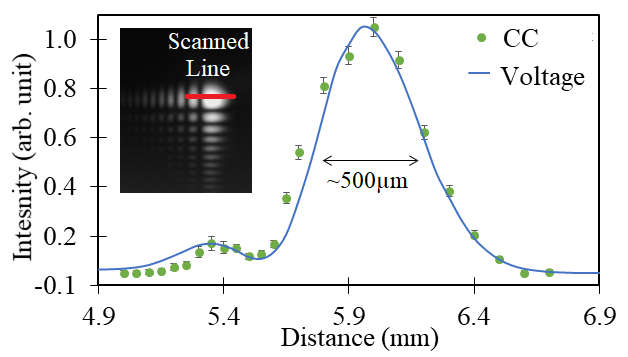}
   \end{center}
    \caption{Line scanning measurement of the Airy beam (blue curve) and Airy photons (green dots). For the latter, the error bars are given as the square root of the coincidence count at each point. Inset shows a CCD image of an Airy Beam, with the red line indicating the scanned area.}
    \label{fig3}
 \end{figure}

We next perform a line scan over the optimized classical Airy beam to examine its transverse mode features, as shown in the Inset of Fig.~\ref{fig3}. A pinhole (diameter $\sim$ 0.25 mm) is placed on a translational stage very close to the fiber coupler. This pinhole is translated across a 2-mm range with a 50 $\mu$m step size, covering the Airy beam’s main lobe and the first side lobe. The intensity of the higher order lobes is too low for this scanning measurement. The result, which uses a a CW laser at 1568.7 nm and an InGaAs switchable gain amplified detector, is plotted as a blue line in Fig.~\ref{fig3}, which measures the main lobe of the Airy beam to be $ \sim 500\, \mu$m FWHM, close to the $ \sim 493\, \mu$m value derived from NIR-IR camera image (Fig. \ref{fig3}, Inset) and the theoretical value of $ \sim 434\, \mu$m. Afterwards, the signal photons are created in the Airy spatial wavefunction, whose profile is similarly scanned and measured by coincident detection with the idler photons. In this measurement, at each scanning point the coincident counts are recorded with pump pulse peak power of 0.2 W for a 600-second integration time. The result is plotted as green dots in Fig. \ref{fig3}, which follows closely the profile of the classical Airy beam, thus verifying the Airy wavefunction of the signal photons. 

\begin{figure}[htbp]
   \begin{center}
 \includegraphics[width=3.5 in]{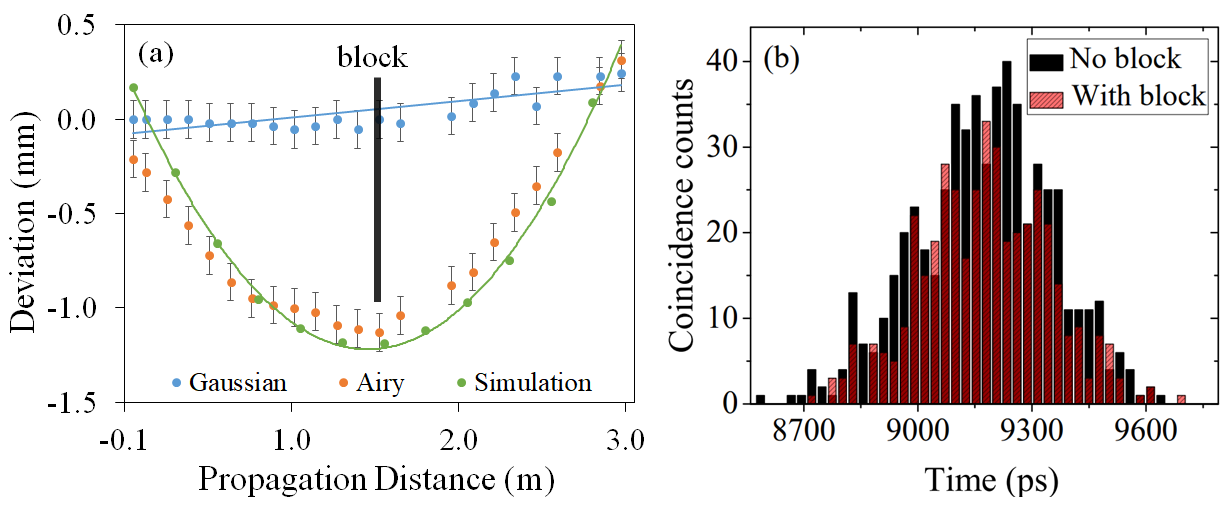}
   \end{center}
    \caption{(a) Peak-intensity trajectories of the Gaussian beam (blue dots) and Airy beam (orange dots), where the error bars correspond to the fluctuations in each's CCD images. The blue line is a linear fit to the Gaussian beam's trajectory and green line is the the simulated trajectory.   Also shown is the block's position. (b) Histogram of the coincident counting between the idler and Airy signal photons without and without the block.}
    \label{fig4}
 \end{figure}

Lastly, we verify the accelerating propagation of the Airy photons. To this end, we first construct a classical Airy beam from a CW laser at 1568.7 nm as in Fig.~\ref{fig1}, and examine its ability to go around an obstacle. As a reference, a collimated Gaussian beam is introduced to overlap with the Airy beam's path right at the focal point of the Fourier lens and after propagating 3 m, where the two are coupled into the same fiber coupler. Figure~\ref{fig4} (a) shows the measured trajectories of the two beams using the NIR-IR camera, along with the simulated position of the Airy beam's main lobe using the exact phase mask for the SLM. The good agreement between the simulation and measurement validates our entire setup. To further examine the parabolic trajectory, a 1-cm wide block is inserted into the path of Gaussian beam by 1.2 mm in the middle between the iris and fiber collimator. The block impedes the Gaussian beam, causing its power collected in the fiber to drop by 90\%. In contrast, it allow the Airy beam to travel around due to the parabolic propagation, causing only a 30\% drop. 

Then we swap in the Airy photons to test their parabolic propagation. Figure \ref{fig4} (b) shows a typical histogram of the coincidence counts between the Airy and idler photons, comparing the cases with and without the block. In this measurement, the pump pulse peak power is 33 mW and the integration time is 60 minutes. As shown, the coincidence counts for the two cases exhibit similar profiles (with shot-noise fluctuations in each time bin), but with the total coincident counts dropping by 28$\%$ when the blocked is inserted, which agrees with the results for the Airy beam. The CAR values without and with the block are 70$\pm 3$ and 62$\pm 4$, respectively. This highlights the potential of steering Airy photons on a controlled trajectory, which can be reconfigured in real time by using a computer programmed SLM without any moving optics. This feature could be exploited for robust free-space quantum communications beyond line-of-sight \cite{YhuOL2010}. 


In conclusion, we have created quantum correlated photons in a spatial Airy-shape wavefunction using spontaneous parametric down conversion and cubic spatial phase modulation. Based on two-photon coincidence measurement, we have found that the spatial modulation on the photons does not introduce any observable degradation to the quantum states of the correlated photon pairs. Contrasting with a collimated Gaussian beam, we have also verified the parabolic propagation trajectory of the Airy photons. Our results suggest the feasibility of using structured photons with exotic spatial and temporal modes \cite{QOAWG2014,QPMS2017} for practical advantages in long distance quantum communication \cite{Tyler09,Nelson14}, quantum imaging \cite{Yao2006,Fickler2013,Patarroyo2013PRL,Chrapkiewicz2016}, quantum key distribution \cite{Etcheverry13}, light-sensitive biological applications  \cite{TaylorBowenPhysRep2016}, and so on. Beside the Airy modes, there exist a multitude of spatiotemproal modulations \cite{Abdollahpour10,Zhang13,Diebel15,Wiersma15} that can be similarly applied to quantum photonic signals to attain distinct advantages for various applications. Finally, the present experiment applies phase modulation directly on the generated photons, which induces loss and significantly reduces the photon production rate. This problem can be solved by using the lossless photon shaping technique demonstrated in \cite{LosslessShaping2011}, which could be a subject of research in the future.

\section*{Acknowledgments}
This research was supported in part by the Office of Naval Research (Award No. N00014-15-1-2393). YPH would like to thank Jianming Wen for initial discussions that motivated this research.




\begin{thebibliography}{1}
\bibitem{Betzig92} E. Betzig and J. K. Trautman, "Near-field optics: microscopy, spectroscopy, and surface modification beyond the diffraction limit,” Science {\bfseries 257}, 189--195 (1992).

\bibitem{Rust06}  M. J. Rust, M. Bates, and X. Zhuang, "Sub-diffraction-limit imaging by stochastic optical reconstruction microscopy (STORM),” Nat. Methods {\bfseries 3}, 793-796 (2006).

\bibitem{Choi11}  Y. Choi, T. D. Yang, C. Fang-Yen, P. Kang, K. J. Lee, R. R. Dasari, M. S. Feld, and W. Choi, "Overcoming the diffraction limit using multiple light scattering in a highly disordered medium,” Phys. Rev. Lett. {\bfseries 107}, 023902 (2011).

\bibitem{Liu17} C. Liu , J. Liu, L. Niu, X. Wei, K. Wang and Z. Yang, "Terahertz circular Airy vortex beams," Sci. Rep. {\bfseries 7}, 3891 (2017).

\bibitem{Berry79} M. V. Berry, N. L. Balazs, "Nonspreading wave packets," Am. J. Phys. {\bfseries 47}, 264 (1979).

\bibitem{Durnin87a} J. Durnin, "Exact solutions for nondiffracting beams. I. The scalar theory," J. Opt. Soc. Am. A {\bfseries 4}, 651-654 (1987).

\bibitem{Durnin87b} J. Durnin, J. J. Miceli, and J. H. Eberly, "Diffraction free beams," Phys. Rev. Lett. {\bfseries 58}, 1499 (1987).

\bibitem{Christodoulides07} G. A. Siviloglou, J. Broky, A. Dogariu, D. N. Christodoulides, "Observation of Accelerating Airy Beams," Phys. Rev. Lett. {\bfseries 99}, 213901 (2007).

\bibitem{Siviloglou07} G. A. Siviloglou, D. N. Christodoulides, "Accelerating finite energy Airy beams," Opt. Lett. {\bfseries 32}, 979 (2007).

\bibitem{Siviloglou08} G. A. Siviloglou, J. Broky, A. Dogariu, D. N. Christodoulides, "Ballistic dynamics of Airy beams," Opt. Lett. {\bfseries 33}, 207 (2008).

\bibitem{Hu10} Y. Hu, P. Zhang, C. Lou, S. Huang, J. Xu, and Z. Chen, "Optimal control of the ballistic motion of Airy beams," 
Opt. Lett. {\bfseries 35}, 2260–2262 (2010).

\bibitem{Hu12} Y. Hu, Z. Chen,  and R. Morandotti, "Self-Accelerating Airy Beams: Generation, Control, and Applications. In Nonlinear Photonics and Novel Optical Phenomena," eds. 1-46 (Springer, New York, 2012).
 
\bibitem{Longhi11} S. Longhi, "Airy beams from a microchip laser", Opt. Lett. {\bfseries 36}, 716-718 (2011).

\bibitem{Porat11} G. Porat, I. Dolev, O. Barlev, and A. Arie, "Airy beam laser," Opt. Lett. {\bfseries 36}, 4119–4121 (2011).

\bibitem{Ellenbogen09} T. Ellenbogen, N. V.Bloch, A. G.-Padowicz, and A. Arie, "Nonlinear generation and manipulation of Airy beams," Nat. Photon. {\bfseries 3}, 395–398 (2009).

\bibitem{Voloch13} N. Voloch-Bloch, Y. Lereah, Y. Lilach, A. Gover, and A. Arie, "Generation of electron Airy beams," Nature {\bfseries 494}, 331-335 (2013).

\bibitem{Baumgart08} J. Baumgart, M. Michael, and K. Dholakia, "Optically mediated particle clearing using Airy wave packets,"  Nat. Photon. {\bfseries 2}, 675-678 (2008).

\bibitem{Singh17} B. K. Singh, H. Nagar, Y. Roichman and A. Arie, "Particle manipulation beyond the diffraction limit using structured super-oscillating light beams," Light: Science and Applications {\bfseries 6}, e17050 (2017).

\bibitem{Minovich11} A. Minovich, A. E. Klein, N. Janunts, T. Pertsch, D. N. Neshev, and Y. S. Kivshar, "Generation and near-field imaging of Airy surface plasmons," Phys. Rev. Lett. { \bfseries 107}, 116802 (2011).

\bibitem{Zhang11} P. Zhang, S. Wang, Y. Liu, X. Yin, C. Lu, Z. Chen, and X. Zhang, "Plasmonic Airy beams with dynamically controlled trajectories," Opt. Lett. {\bfseries 36}, 3191-3193 (2011).

\bibitem{Salandrino10} A. Salandrino, and D. N. Christodoulides, "Airy plasmon: a nondiffracting surface wave," Opt. Lett. {\bfseries 35}, 2082–2084 (2010).

\bibitem{Chong10} A. Chong, W. Renninger, D. N. Christodoulides, and F. W. Wise, "Airy–Bessel wave packets as versatile linear light bullets," Nat. Photon. {\bfseries 4}, 103 (2010).

\bibitem{Abdollahpour10} D. Abdollahpour, S. Suntsov, D. G. Papazoglou, and S. Tzortzakis, "Spatiotemporal airy light bullets in the linear and nonlinear regimes," Phys. Rev. Lett. {\bfseries 105}, 253901 (2010).

\bibitem{Panagiotopoulos13} P. Panagiotopoulos, D.G. Papazoglou, A. Couairon and S. Tzortzakis, "Sharply autofocused ring-Airy beams transforming into non-linear intense light bullets," Nat. Comm. {\bfseries 4}, 2622 (2013).

\bibitem{Polynkin09} P. Polynkin, M. Kolesik, J. V. Moloney, G. A. Siviloglou, and D. N. Christodoulides, "Curved plasma channel generation using ultra intense Airy beams," Science {\bfseries 324}, 229-232 (2009).

\bibitem{Mathis12} A. Mathis, F. Courvoisiera, L. Froehly, L. Furfaro, M. Jacquot, P. A. Lacourt, and J. M. Dudley, "Micromachining along a curve: Femtosecond laser micromachining of curved profiles in diamond and silicon using accelerating beams," Appl. Phys. Lett. {\bfseries 101}, 071110 (2012).

\bibitem{Jia14} S. Jia, J. C. Vaughan, and X. Zhuang, "Isotropic 3D Super-resolution Imaging with a Self-bending Point Spread Function," Nat. Photon. {\bfseries 8}, 302-306 (2014).

\bibitem{Aadhi16} A. Aadhi, N. A. Chaitanya, M. V. Jabir, P. Vaity, R. P. Singh and G. K. Samanta, "Airy beam optical parametric oscillator," Sci. Rep. {\bfseries 6}, 25245 (2016).

\bibitem{Aadhi17} A. Aadhi, V. Sharma, N. A. Chaitanya and G. K. Samanta, "Multi-gigahertz, femtosecond Airy beam optical parametric oscillator pumped at 78 MHz," Sci. Rep. {\bfseries 7}, 43913 (2017).

\bibitem{Wei14} D. Wei, Y. Yu, M. Cao, L. Zhang, F. Ye, W. Guo, S. Zhang, H. Gao, and F. Li, "Generation of Airy beams by four-wave mixing in Rubidium vapor cell," Opt. Lett. {\bfseries 39}, 4557-4560 (2014).

\bibitem{Vettenburg14} T. Vettenburg, H. I. C. Dalgarno, J. Nylk, C. Coll-Lladó, D. E K Ferrier, T. Čižmár,	F. J. Gunn-Moore	and K. Dholakia, "Light-sheet microscopy using an Airy beam," Nature Methods {\bfseries 11}, 541–544 (2014).

\bibitem{Extance15}  A. Extance, "Military technology: Laser weapons get real," Nature {\bfseries 521}, 408–410 (2015).

\bibitem{Gu10} Y. Gu and G. Gbur, "Scintillation of Airy beam arrays in atmospheric turbulence," Opt. Lett. {\bfseries 35}, 3456-3458 (2010).

\bibitem{Rose13} P. Rose, F. Diebel, M. Boguslawski, and C. Denz, "Airy Beam Induced Optical Routing,"  Opt. Photon. News {\bfseries 24}, 45 (2013).

\bibitem{Rose13b} P. Rose, F. Diebel, M. Boguslawski, and C. Denz, "Airy beam induced optical routing," Appl. Phys. Lett. {\bfseries 102}, 101101 (2013).

\bibitem{KILLINGER} D. K. Killinger and N. Menyuk, "Laser Remote Sensing of the Atmosphere," Science {\bfseries 235} (4784), 37-45 (1987).

\bibitem{Kirmani14} A. Kirmani, D. Venkatraman, D. Shin, A. Colaço, F. N. Wong, J.H. Shapiro, V.K. Goyal, "First-photon imaging," Science {\bfseries 343}, 58-61 (2014). 

\bibitem{Maccarone15} A. Maccarone, A. McCarthy, X. Ren, R.E. Warburton, A.M. Wallace, J. Moffat, Y. Petillot, and G.S. Buller, "Underwater depth imaging using time–correlated single–photon counting," Opt. Exp. 23, 33911–33926 (2015).


\bibitem{Piksarv17} P. Piksarv, D. Marti, T. Le, A. Unterhuber, L. H. Forbes, M. R. Andrews, A. Sting, W. Drexler, P. E. Andersen and K. Dholakia, "Integrated single- and two-photon light sheet microscopy using accelerating beams," Sci. Rep. {\bfseries 7}, 1435 (2017).

\bibitem{Giovannini15} D. Giovannini, J. Romero,  V. Potoček, G. Ferenczi, F. Speirits, S. M. Barnett, D. Faccio, M. J. Padgett, "Optics. Spatially structured photons that travel in free space slower than the speed of light," Science 347, 857 (2015).

\bibitem{Villa12} F. Villa, D. Bronzi, S. Bellisai, G. Boso, A. B. Shehata, C. Scarcella, A. Tosi, F. Zappa, S. Tisa, D. Durini, S. Weyers, W. Brockherde, "SPAD imagers for remote sensing at the single-photon level," Proc. SPIE 8542, Electro-Optical Remote Sensing, Photonic Technologies, and Applications VI, 85420G (2012).

\bibitem{Tyler09} G. A. Tyler and R. W. Boyd, "Influence of atmospheric turbulence on the propagation of quantum states of light carrying orbital angular momentum," Opt. Lett. {\bfseries 34}, 142-144 (2009).

\bibitem{Nelson14} W. Nelson, J. P. Palastro, C. C. Davis, and P. Sprangle, "Propagation of Bessel and Airy beams through atmospheric turbulence," J. Opt. Soc. Am. A {\bfseries 31}, 603-609 (2014).


\bibitem{Yao2006} E. Yao, S. F.-Arnold, J. Courtial, M. J. Padgett, and S. M. Barnett, "Observation of quantum entanglement using spatial light modulators," Opt. Exp. {\bfseries 14}, 13089-13094 (2006).

\bibitem{Fickler2013} R. Fickler, M. Krenn, R. Lapkiewicz, S. Ramelow, and A. Zeilinger, “Real-time imaging of quantum entanglement,” Sci. Rep. {\bfseries 3}, 1914 (2013).

\bibitem{Patarroyo2013PRL} N. U.-Patarroyo, A. Fraine, D. S. Simon, O. Minaeva, and A. V. Sergienko, "Object identification using correlated orbital angular momentum states," Phys. Rev. Lett. {\bfseries 110}, 043601 (2013).

\bibitem{Chrapkiewicz2016} 
R. Chrapkiewicz, M. Jachura, K. Banaszek and W. Wasilewski, "Hologram of a single photon," Nat. Photon. {\bfseries 10}, 576--579 (2016).

\bibitem{Etcheverry13} S. Etcheverry, G. Canas, E. S. Gomez, W. A. T. Nogueira, C. Saavedra, G. B. Xavier, and G Lima, "Quantum key distribution session with 16-dimensional photonic states," Sci. Rep. {\bfseries 3}, 2316, (2013).

\bibitem{TaylorBowenPhysRep2016} M. A. Taylor and W. P. Bowen, "Quantum metrology and its application in biology," Phys. Rep. {\bfseries 615}, 1-59, (2016).

\bibitem{Ghosh87} R. Ghosh, and L. Mandel, "Observation of Nonclassical Effects in the Interference of Two Photons," Phys. Rev. Lett.,{\bfseries 59}, 1903-1905 (1987).

\bibitem{Takesue05} H. Takesue, K. Inoue, O. Tadanaga, Y. Nishida, and M. Asobe, "Generation of pulsed polarization-entangled photon pairs in a 1.55-µm band with a periodically poled lithium niobate waveguide and an orthogonal polarization delay circuit," Opt. Lett. {\bfseries 30}, 293-295 (2005).

\bibitem{Ma09} L. Ma, O. Slattery, T. Chang, and X. Tang, "Non-degenerated sequential time-bin entanglement generation using periodically poled ktp waveguide," Opt. Exp. {\bfseries 17}, 15799–15807, (2009).

\bibitem{Oza14} N. N. Oza, Y.-P. Huang, and P. Kumar, "Entanglement-Preserving Photonic Switching: Full Cross-Bar Operation With Quantum Data Streams," IEEE Photon. Tech. Lett. {\bfseries 26}, 4 (2014).

\bibitem{YuPing14} K.-Y. Wang, V. G. Velev, K. F. Lee, A. S. Kowligy, P. Kumar, M. A. Foster, A. C. Foster, and Y.-P. Huang, "Multichannel photon-pair generation using hydrogenated amorphous silicon waveguides," Opt. Lett. {\bfseries 39}, 914-917 (2014).

\bibitem{XZWang12} X.-Z Wang, Q. Li, and Q. Wang, "Arbitrary scanning of the Airy beams using additional phase grating with cubic phase mask," Appl. Opt. {\bfseries 51}, 6726 (2012).

\bibitem{YhuOL2010} Y. Hu, P. Zhang, C. Lou, S. Huang, J. Xu, and Z. Chen, "Optimal control of the ballistic motion of Airy beams," Opt. Lett. {\bfseries 35}, 2260-2262 (2010).

\bibitem{QOAWG2014} A. S. Kowligy, P. Manurkar, N. V. Corzo, V. G. Velev, M. Silver, R. P. Scott, S. J. B. Yoo, P. Kumar, G. S. Kanter and Y.-P. Huang, "Quantum optical arbitrary waveform manipulation and measurement in real time," Opt. Express {\bfseries22}, 27942 (2014). 

\bibitem{QPMS2017} A. Shahverdi, Y. M. Sua, L. Tumeh, and Y.-P. Huang, "Quantum Parametric Mode Sorting: Beating the Time-Frequency Filtering," Scientific Report {\bfseries 7} 6495 (2017). 

\bibitem{Zhang13} Y. Zhang, M. Belic, Z. Wu, H. Zheng, K. Lu, Y. Li, and Y. Zhang, "Soliton pair generation in the interactions of Airy and nonlinear accelerating beams," Opt. Lett. {\bfseries 38}, 4585-4588 (2013).

\bibitem{Wiersma15} N. Wiersma, N. Marsal, M. Sciamanna and D. Wolfersberger, "Spatiotemporal dynamics of counterpropagating Airy beams," Sci. Rep. {\bfseries 5}, 13463 (2015).

\bibitem{Diebel15} F. Diebel, B. M. Bokić, D.V. Timotijevic, D.M.J. Savic, and C. Denz, "Soliton formation by decelerating interacting Airy beams," Opt. Exp. {\bfseries 23}, 24351-24361 (2015).

\bibitem{LosslessShaping2011}K. G. K{\"o}pr{\"u}l{\"u}, Y.-P. Huang, G. A. Barbosa, and P. Kumar, "Lossless single-photon shaping via heralding," Opt. Lett. {\bfseries 36}, 1674 (2011). 

\end{thebibliography}
\end{document}